\documentclass[aps,reprint,longbibliography,superscriptaddress]{revtex4-2}
\usepackage{mathrsfs}
\usepackage{amsmath,gensymb}
\usepackage{amsfonts}
\usepackage{amssymb}
\usepackage{amsthm}
\usepackage{graphicx}
\usepackage{natbib}
\usepackage{color}
\usepackage{hyperref}
\usepackage{bm}
\usepackage[caption=false]{subfig}
\usepackage{verbatim}
\usepackage[normalem]{ulem}
\usepackage{soul}

\begin{document}
\title{Proximity effect and $p$-wave superconductivity in $s$-wave superconductor/helimagnet heterostructures}

\author{G. A. Bobkov}
\affiliation{Moscow Institute of Physics and Technology, Dolgoprudny, 141700 Moscow region, Russia}

\author{A. V. Kornev}
\affiliation{Moscow Institute of Physics and Technology, Dolgoprudny, 141700 Moscow region, Russia}

\author{A. M. Bobkov}
\affiliation{Moscow Institute of Physics and Technology, Dolgoprudny, 141700 Moscow region, Russia}

\author{I. V. Bobkova}
\affiliation{Moscow Institute of Physics and Technology, Dolgoprudny, 141700 Moscow region, Russia}
\affiliation{National Research University Higher School of Economics, 101000 Moscow, Russia}

\begin{abstract}
It is known that in contrast to homogeneous ferromagnetism helical magnetism is compatible with superconductivity and has only weak suppressive effect on superconducting critical temperature. Despite this fact it induces $p$-wave triplet superconducting correlations in homogeneous superconducting systems with intrinsic helical magnetism. The combination of these two facts indicates a high potential for the application of such systems in disspationless spintronics.  For this reason,  we investigate the proximity effect in atomically thin superconductor/helical (conical) magnet heterostructures (SC/HM). It is shown that in SC/HM heterostructures the strength of the proximity effect and, in particular, amplitude of $p$-wave triplet superconductivity and the degree of superconductivity suppression are complex functions of the magnet exchange field and filling factors of the magnet and the superconductor. Further we demonstrate that $p$-wave correlations ensure transport spin supercurrent flow in the SC/HM heterostructure with conical magnets and unveil the physical relationship between the transport spin supercurrent, degree of  magnet conicity and internal structure of $p$-wave correlations in  momentum space.
\end{abstract}

\maketitle

\section{Introduction}

The interplay between superconductivity and magnetism has been a topic of ongoing research for a long time. Uniform ferromagnetism
is known to be strongly antagonistic to singlet superconductivity,
due to the suppression by the orbital effect of the stray magnetic field and also due to the exchange field that tends to breaks the singlet Cooper pairs \cite{Buzdin2005}. However, in most cases if the magnetization is spatially non-uniform on a scale smaller than the superconducting
coherence length, both of these effects are dramatically
suppressed. The known exceptions are superconducting systems with  staggered antiferromagnetic order \cite{Bobkov2022,Bobkov2023} and with altermagnetic order \cite{Vasiakin2025}, where the stray fields are absent due to negligible net magnetization, but the spatially-inhomogeneous or anisotropic nonlocal exchange field breaks singlet Cooper pairs. However, it was reported in the literature that the helical  magnetic state is much less destructive to superconductivity \cite{Sukhachov2025}.   In fact, the onset of superconductivity itself 
can drive a uniform ferromagnet into such a non-uniform state \cite{Anderson1959,Bulaevskii1985,Bergeret2000,Neupert2011}. For example, helical magnetism coexists with superconductivity
in some materials,  where magnetism originates from the partially
filled $f$-orbitals of rare earth atoms, as is the case for the
compounds such as $\mathrm {HoMo_6S_8}$, $\mathrm {ErRh_4B_4}$, $\mathrm {TmNi_2B_2C_{17}}$ \cite{Bennemann}.

The Hamiltonian of the conductivity electrons in helimagnetic systems (HM) contains a rotating exchange field, which is caused by the localized spin moments. The conducting system with such a helimagnetic exchange field pattern  is equivalent to an electron system with $p$-wave magnetic texture  or a persistent magnetic helix \cite{Schliemann2003,Bernevig2006,Schliemann2017} in the presence of an external Zeeman field \cite{Martin2012}. Materials and systems with $p$-wave magnetic textures and helical magnetic textures are currently being actively studied, primarily due to the various magnetoelectric effects that have been predicted for them. Among them are possibility of a transport spin current induced by an applied electric field \cite{Brekke2024}, transverse spin current in junctions with normal
metal \cite{Hedayati2025}, non-relativistic Edelstein effect \cite{Chakraborty2025,Yu2025} and spin-galvanic effect \cite{Kokkeler2025}. 

For hybrid superconductor/$p$-wave magnet systems, even-frequency spin-triplet $p$-wave correlations were predicted to be induced \cite{Maeda2025}. Because the spin helix plays a role analogous to the combination of the spin-orbit interaction and of the exchange field, in superconducting systems it can lead to formation
of zero-energy Majorana surface states \cite{Martin2012}. $p$-wave
magnetism affects Andreev reflections \cite{Maeda2024} and the Josephson effect \cite{Fukaya2025}. A number of magnetoelectric and spintronic effects specific for superconducting systems with helical spin textures or $p$-wave magnets were also reported. For example, the superconducting spin-galvanic effect was predicted for $p$-wave magnets in the presence of the applied magnetic field \cite{Kokkeler2025} and in superconducting heterostructures with helical magnets \cite{Rabinovich2018,Rabinovich2019,Meng2019}. Also, it was shown that $p$-wave magnets allow for transverse spin conductance, which is further enhanced by Andreev reflections \cite{Sukhachov2025}.

Thus, superconducting heterostructures with helical spin textures and $p$-wave magnets demonstrate rich physics and serious potential for low-dissipation spintronics \cite{Linder2015,Eschrig2015}. Up to now for theoretical studies of thin-film heterostructures consisting of an unconventional magnet and a superconductor, a simplified model is used, within the framework of which the heterostructure is considered as a homogeneous superconductor with induced magnetism or, equivalently, as a homogeneous unconventional magnet in which the superconducting order parameter is induced using the proximity effect with a conventional superconductor \cite{Sukhachov2025,Maeda2025,Fu2025} (SC+HM model for brevity). In this paper, we investigate to what extent this approach is adequate for describing thin-film superconductor/helical magnet (SC/HM) heterostructures. Of particular relevance is the question of the behavior and magnitude of $p$-wave triplet superconducting correlations that can be induced as a result of the proximity effect in the heterostructure in comparison with the SC+HM model. The magnitude of $p$-wave superconducting correlations, their internal structure and dependence on the system parameters are of interest to us mainly due to the fact that, as we predict in this paper, they carry a non-dissipative transport spin current in such structures.

In spite of the active research of superconducting unconventional magnets and superconductor/$p$-wave magnet heterostructures, as discussed above, the issue of non-dissipative spin transport in such systems has not yet received due attention. At the same time materials and heterostructures with a controllable capability to transfer spin currents over long distances are promising for next-generation all-electrical spintronic science and technology \cite{Shrivastava2021,Geim2013}. Among them the dissipationless spin current attract special attention. A significant number of key studies focus on spin currents in S/F heterostructures \cite{Grein2009,Jeon2020,Brydon2013,Dahir2022,Ojajarvi2022,Ouassou2017,Ouassou2019,Alidoust2010,Jacobsen2016,Gomperud2015,Brydon2011,Linder2017,Bobkova2017,Bobkova2018,Aunsmo2024}. However, the majority primarily examine spin supercurrents, which are present only in restricted spatial areas, such as the weak links of S/F/S Josephson junctions or regions approximately equal to the superconducting coherence length near the interfaces of S/F and F/S/F structures. The primary interest in these spin supercurrents lies in their ability to generate spin transfer torques and influence the magnetization dynamics of ferromagnetic components \cite{Zhu2004,Zhu2005,Holmqvist2011,Linder2011,Halterman2016,Kulagina2014,Linder2012,Takashima2017}. Until now the superconducting spin currents providing spin transport over long distances, have been discussed mainly in relation to systems with spin-orbit interaction \cite{He2019,Bobkov2024}, $\mathrm{^3He}$ \cite{Leggett1975}, triplet superconductors \cite{Asano2005,Asano2006} and noncentrosymmetric superconductors \cite{Leurs2008,He2019,Bergeret2016}. Here we expand this important class of materials demonstrating that the dissipationless spin transport over long distances  can also be carried by proximity-induced $p$-wave superconducting Cooper pairs in SC/HM thin-film heterostructures with {\it conventional  singlet $s$-wave superconductors in the absence of spin-orbit coupling}. Here ``spin transport over long distances" means that the spin current is conserved due to the conservation of the electric supercurrent, which provides the spinful electron flow in this case. We analyze possible magnitude and dependence of this spin current on the internal structure of the magnetic texture (helical or conical).  

The paper is organized as follows. Sec.~\ref{sec:model} describes the system under consideration and the method used to calculate the superconducting correlations and spin current. In Sec.~\ref{sec:op} we compare the proximity-induced order parameter suppression and other important features of the proximity effect in SC/HM heterostructures and SC+HM homogeneous model. Sec.~\ref{sec:triplets} discusses the structure of $p$-wave triplet correlations. Sec.~\ref{sec:current} presents the results of the superconducting spin current calculation, and in Sec.~\ref{sec:conclusions} our conclusions are formulated.

\section{Model and method}
\label{sec:model}

\begin{figure}[t]
\includegraphics[width=0.80\columnwidth]{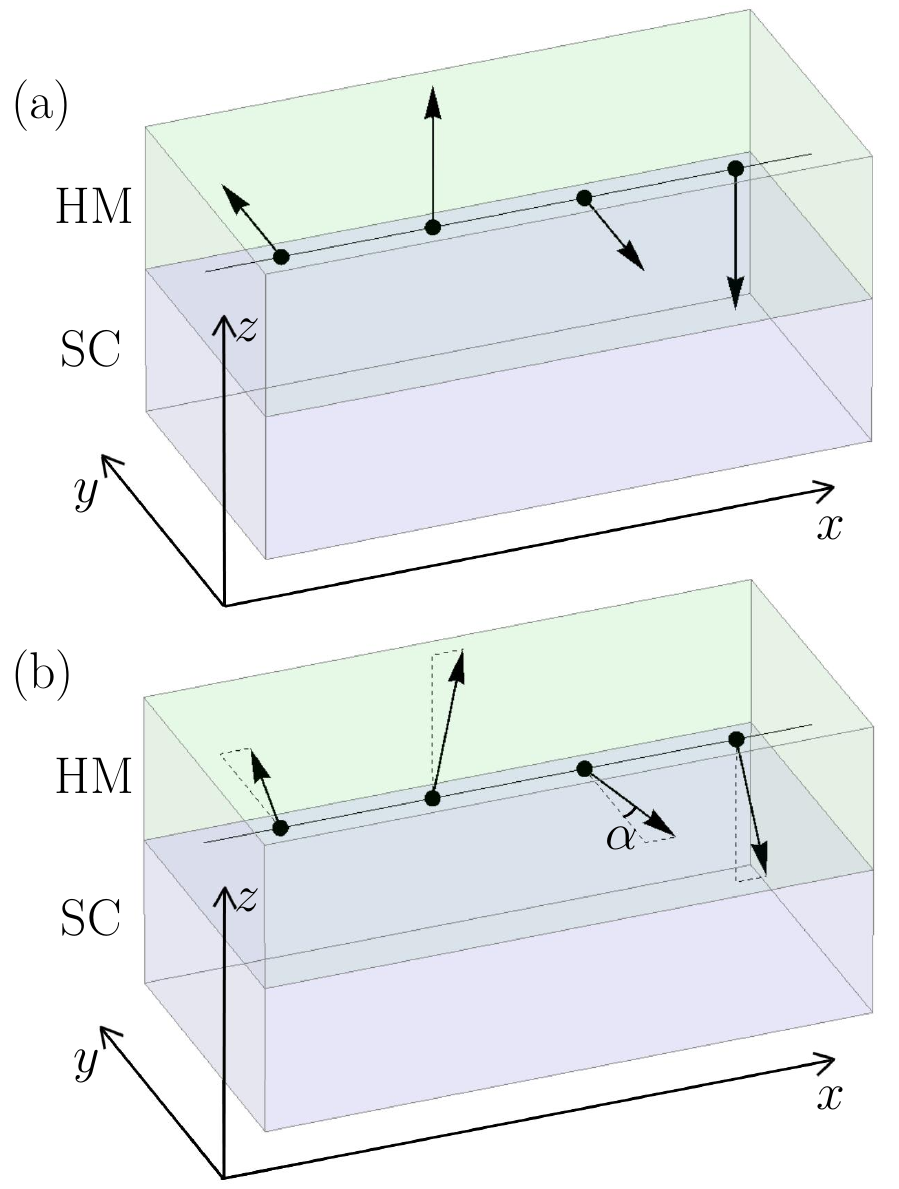}
\caption{Sketch of the superconductor/helical magnet (SC/HM) bilayer (a) and the superconductor/conical magnet bilayer (b). The SC and HM layers are assumed to be 2D. The spatial distribution of the magnetization in one magnetic unit cell, containing 4 sites, is shown by black arrows. The HM magnetization is assumed to be homogeneous along the $y$-direction. } 
 \label{fig:sketch}
\end{figure}

System under consideration is shown in Fig.~\ref{fig:sketch}. It
consists of two 2D layers: a 2D magnetic metal (HM) and a 2D superconductor (SC). The magnet has helical [Fig.~\ref{fig:sketch}(a)] or conical [Fig.~\ref{fig:sketch}(b)] magnetization resulting in the helical or conical exchange field acting on the conductivity electrons in the magnet. To model this system we consider a lattice model with a helical
magnetic texture and a mean-field on-site superconductivity
described by the tight-binding Hamiltonian
\begin{widetext}
\begin{align}
					\hat{H} = 
			-\sum_{{\langle {i},{j};\nu,\bar{\nu}\rangle\sigma}} \hat{c}_{i\nu\sigma}^{\dagger}
			\begin{pmatrix}
				t_S & 0 \\
				0 & t_M
			\end{pmatrix}
			\hat{c}_{j\bar{\nu}\sigma}
			-  
			\sum_{i,\nu;\sigma} \hat{c}_{i\nu\sigma}^{\dagger} 
			\begin{pmatrix}
				0 & t_{SM} \\
				t_{SM} & 0
			\end{pmatrix}
			\hat{c}_{i\nu\sigma}
			-
			\sum_{i,\nu;\sigma} \hat{c}_{i\nu\sigma}^{\dagger} 
			\begin{pmatrix}
				\mu_S & 0 \\
				0 & \mu_M
			\end{pmatrix}
			\hat{c}_{i\nu\sigma} \nonumber \\+
			\sum_{i,\nu} 
			\left(
			\hat{c}_{i\nu\uparrow}
			\begin{pmatrix}
				\Delta_{i\nu} & 0 \\
				0 & 0
			\end{pmatrix}
			\hat{c}_{i\nu\downarrow}
			+
			\hat{c}_{i\nu\downarrow}^{\dagger}
			\begin{pmatrix}
				\Delta^*_{i\nu} & 0 \\
				0 & 0
			\end{pmatrix}
			\hat{c}_{i\nu\uparrow}^{\dagger}
			\right) 	 
			+ \sum_{i,\nu;\alpha\beta} \hat{c}_{i\nu\alpha}^{\dagger} 
			\begin{pmatrix}
				0 & 0 \\
				0 & (\bm {h}_{i\nu}\bm{\sigma})_{\alpha\beta}
			\end{pmatrix}
			\hat{c}_{i\nu\beta},
            \label{eq:ham}
\end{align}
\end{widetext}
where the exchange field unit vector $\bm h_{i \nu} = \bm h_x + \bm h_{\perp \nu}$ rotates in a plane
perpendicular to the helical axis $x$. $\bm h_x$ is the component along the $x$-axis, which is assumed to be homogeneous in space. We assume that the angle between the directions of the rotating component $\bm h_{\perp \nu}$ at the nearest neighbor sites along the $x$-axis is $\pi/2$. Therefore, we consider a commensurate spin helix with a period equal to $4$ atomic sites. The model is extended
to two dimensions by making infinite number of copies of the described helical chain stacked in the $y$-direction.  Accordingly, we introduce a $4 \times 1$ unit cell, see Fig.~\ref{fig:sketch}. Index $i$ numbers cells, and index $\nu = \overline{1,4}$ numbers sites within a cell (sublattice index). 
Vector	$\hat{c}_{i\nu\sigma} = 
		\begin{pmatrix}
			\hat{c}_{i \nu \sigma}^{S}, &
			\hat{c}_{i \nu \sigma}^{M}
		\end{pmatrix}^T$ consists of the electron annihilation operators $\hat c_{i \sigma \nu}^S$ and $\hat c_{i \sigma \nu}^M$ for an electron at site $i \nu$ in the SC and HM layers, respectively, 
with spin projection $\sigma = \uparrow, \downarrow$ on the $z$-axis. The parameter $t_{S(M)}$ corresponds to hopping between nearest neighbors $\langle i, j; \nu, \nu'\rangle$ in the SC (HM) layer and $\mu_{S(M)}$ is the on-site energy of the SC and HM layers, respectively, which determines the filling factors of the conduction band of the materials and in case of isolated layers corresponds to their chemical potentials. $t_{SM}$ describes hopping between SC and HM layers. The superconducting
order parameter $\Delta_{i \nu}$ is calculated self-consistently according to $\Delta_{i \nu} = \lambda \langle \hat c_{i \nu \downarrow} \hat c_{i \nu \uparrow} \rangle$, assuming a constant attractive pairing potential $\lambda$ in the system. Then we apply a supercurrent $\bm j_e$ along the SC layer and study the spin supercurrent generated by this charge supercurrent.

For the calculation of the amplitudes of triplet Cooper pair correlations and the spin current we use the Green's functions technique. The Matsubara Green's function is $32 \times 32$ matrix in the direct product of spin space,  particle-hole space, layer space and sublattice space. Introducing the Nambu-sublattice-layer spinor $\check c_i = (\hat c_{i 1 \uparrow}, ..., \hat c_{i 4 \uparrow}, \hat c_{i 1 \downarrow}, ..., \hat c_{i 4 \downarrow},\hat c_{i 1 \uparrow}^\dagger, ..., \hat c_{i 4 \uparrow}^\dagger, \hat c_{i 1 \downarrow}^\dagger, ..., \hat c_{i 4 \downarrow}^\dagger)^T$ we define the Green's function as follows: 
\begin{equation}
		\check{G}_{ij}(\tau) = -  \hat U \langle T_\tau \check{c}_{i}(\tau) \check{c}_{j}^{\dagger}(0) \rangle \hat U
		\label{eq:G_def}
	\end{equation}
where $\langle T_\tau ... \rangle$ means  imaginary time-ordered thermal averaging. Introducing Pauli matrices in spin, particle-hole and layer spaces: $\sigma_k$, $\tau_k$ and $\rho_k$ ($k=0,x,y,z$) the matrix $\hat U$ can be written as $\hat U = \rho_0 [\sigma_0 (\tau_0 + \tau_z)/2 - i \sigma_y (\tau_0 - \tau_z)/2]$. The Green's function (\ref{eq:G_def}) obeys the Gor'kov equation, which can be derived in a standard way by taking the derivative of the Green's function with respect to imaginary time $\tau$:
\begin{equation}
		\frac{\partial \check{G}_{ij}(\tau)}{\partial \tau} 
		= -\langle T_\tau \frac{\partial \check{c}_{i}(\tau)}{\partial  \tau} \check{c}_{j}^\dagger(0) \rangle
		- \delta_{ij}\delta(\tau).
	\end{equation}
Then the resulting Gor'kov equation takes the form:
\begin{align}
		\Bigl(&-\frac{\partial}{\partial \tau}\tau_z + 
						\hat{t}\hat{j}+t_{SM} \rho_x + \hat{\mu}-\rho_\downarrow \tau_z (\check{\bm{h}}_i, {\bm{\sigma}}) + \nonumber \\
 &\rho_\uparrow \tau_z \check{\Delta}_i \Bigr)
			\check{G}_{ij}(\tau)
		 = \delta_{ij}\delta(\tau),
         \label{eq:Gorkov_coordinate}
	\end{align}
where $\rho_{\uparrow,\downarrow} = (\rho_0 \pm \rho_z)/2$, $\hat t = t_S \rho_\uparrow + t_M \rho_\downarrow$, $\hat \mu = \mu_S \rho_\uparrow + \mu_M \rho_\downarrow$. $\check \Delta_i = diag(\hat \Delta_{i1}, \hat \Delta_{i2}, \hat \Delta_{i3}, \hat \Delta_{i4})_l$ is a diagonal matrix in the sublattice space, which is labeled by subscript $l$. $\hat \Delta_{i \nu} = \Delta_{i \nu} \tau_+ + \Delta_{i \nu}^* \tau_-$ with $\tau_\pm = (\tau_x \pm i \tau_y)/2$. 
\begin{align}
(\check {\bm h}_i , \bm \sigma) = diag(\bm h_{i 1} \cdot \bm \sigma, \bm h_{i 2} \cdot \bm \sigma, \bm h_{i 3} \cdot \bm \sigma, \bm h_{i 4} \cdot \bm \sigma)_l, \nonumber \\
\bm h_{i1,3} = h_x \bm e_x \pm h_\perp \bm e_y, ~~ \bm h_{i2,4} = h_x \bm e_x \pm h_\perp \bm e_z .
\label{eq:h}
\end{align}
Operator $\hat j $ is defined as
\begin{align}
&\hat j \check G_{ij}^{1m} = \check G_{i + \bm a_y,j}^{1m} + \check G_{i - \bm a_y,j}^{1m} + \check G_{ij}^{2m} + \check G_{i - 4 \bm a_x,j}^{4m}, \nonumber \\
&\hat j \check G_{ij}^{2m} = \check G_{i + \bm a_y,j}^{2m} + \check G_{i - \bm a_y,j}^{2m} + \check G_{ij}^{3m} + \check G_{i j}^{1m}, \nonumber \\
&\hat j \check G_{ij}^{3m} = \check G_{i + \bm a_y,j}^{3m} + \check G_{i - \bm a_y,j}^{3m} + \check G_{ij}^{4m} + \check G_{ij}^{2m}, \nonumber \\
&\hat j \check G_{ij}^{4m} = \check G_{i + \bm a_y,j}^{4m} + \check G_{i - \bm a_y,j}^{4m} + \check G_{i + 4 \bm a_x,j}^{1m} + \check G_{ij}^{3m} ,
\label{op_j}
\end{align}
where $\check G_{ij}^{km}$ means $km$-element of the matrix $\check G_{ij}$ in the sublattice space, and $\bm a_{x,y}$ are basis vectors of the square lattice in plane of each layer.

In the presence of the supercurrent applied in plane of the SC layer the superconducting order parameter takes the form $\Delta_{i \nu} = \Delta \exp{[i \bm q (\bm i + (\nu-1)\bm a_x)]}$, where $\Delta$ is the absolute value of the superconducting order parameter, which is spatially uniform, and $\bm i$ is the in-plane radius-vector corresponding to the first site of cell $i$. The Cooper pair momentum $\bm q$ is adjusted in such a way that a given supercurrent flows through the system. Then making the unitary transformation 
\begin{align}
&\check G_{ij} = \check V_{\bm i} \check {\tilde G}_{ij} \check V_{\bm j}^\dagger, \nonumber \\
&\check V_{\bm i} = \exp[(i \bm q \tau_z/2)diag(\bm i, \bm i + \bm a_x, \bm i + 2 \bm  a_x, \bm i + 3 \bm a_x)_l],
\label{eq:current_transform}
\end{align}
we can eliminate the spatially-dependent phase of the order parameter. Thus, the Green's function $\check {\tilde G}_{ij}$ only depends on $\bm i - \bm j$ and further we can turn to the momentum representation and to the representation of Matsubara frequencies $\omega_m = \pi T(2m+1)$:
\begin{align}
\check {\tilde G}_{ij}(\tau) = T \sum \limits_{\omega_m} \int \frac{d^2 p}{(2 \pi)^2} \check {\tilde G}(\bm p, \omega_m)e^{-i \omega_m \tau + i \bm p (\bm i- \bm j)}.
\label{eq:Fourier}
\end{align}
Then the Green's function $\check {\tilde G} (\bm p, \omega_m)$ obeys the following equation:
\begin{align}
		\Bigl(
		&i\omega_m\tau_z + 
		t_{SM} \rho_x 
		-\rho_\downarrow \tau_z(\check{\bm{h}},\bm{\sigma}) \nonumber \\
        &+ \rho_\uparrow \tau_z\check{\Delta}
		\Bigr)
		\check{G}(\bm{p},\omega_m) +\Xi(\bm p, \bm q)
		= 1,
        \label{eq:gorkov_mixed}
	\end{align}
where $\check \Delta = diag(\hat \Delta, \hat \Delta, \hat \Delta, \hat \Delta)_l$ with $\hat \Delta = \Delta \tau_x$ is the spatially homogeneous order parameter. The spatially inhomogeneous phase is already eliminated by the unitary transformation Eq.~(\ref{eq:current_transform}) and $(\check {\bm h}, \bm \sigma)$ is expressed by Eq.~(\ref{eq:h}) and also does not depend on $\bm i$. Introducing the explicit structure of the Green's function $\check {\tilde G}(\bm p, \omega_m)$ in the particle-hole space
\begin{align}
    \check {\tilde G}(\bm p, \omega_m)=\left(
\begin{array}{cc}
 G(\bm p, \omega_m) &  F(\bm p, \omega_m) \\
\bar F(\bm p, \omega_m) & \bar G(\bm p, \omega_m)
\end{array}
\right)_\tau,
\label{eq:Green_nambu}
\end{align}
we can write the term $\Xi(\bm p, \bm q)$ as follows:
\begin{align}
    \Xi(\bm p, \bm q)=\left(
\begin{array}{cc}
\hat \xi(\bm p)  G(\bm p)& \hat \xi(\bm p+\bm q/2)  F(\bm p)\\
\hat \xi(\bm p-\bm q/2) \bar F(\bm p)& \hat \xi(\bm p) \bar G(\bm p)
\end{array}
\right)_\tau ,
\label{eq:Xi}
\end{align}
argument $\omega_m$ of the Green's functions is omitted for brevity. The matrix normal state electronic dispersion takes the form:
\begin{align}
\hat \xi(\bm p) = ~~~~~~~~~~~~~~~~~~~~~~~~~~~~~~~~~~~~~
\nonumber \\
-\hat t 
\left(
\begin{array}{cccc}
2 \cos p_y a_y & 1 & 0 & e^{-4 i p_x a_x} \\
1 & 2 \cos p_y a_y & 1 & 0 \\
0 & 1 & 2 \cos p_y a_y & 1 \\
e^{4 i p_x a_x} & 0 & 1 & 2 \cos p_y a_y 
\end{array}
\right) - \hat \mu .
\label{eq:xi}
\end{align}
For an experimental realization of SC/HM heterostructures, where the developed model or its straightforward generalizations can be applied, one can consider, for example, $\rm{EuP_3}$ \cite{Mayo2022,Mayo2025}, $\rm{DyTe_3}$ \cite{Akatsuka2024}, $\rm NiI_2$ \cite{Ju2021,Song2022,Amini2024} or $\rm{Cr_{1/3}NbS_22}$ \cite{Cao2020,Xie2023} as candidate magnetic materials, and approaching the monolayer limit elemental superconducting materials such as $\rm{Pb}$ \cite{Qin2009} or $\rm{Al}$ \cite{Werner2023}. At the same time actively  studied 2D transition-metal dichalcogenides superconductors like $\rm{NbSe_2}$ or $\rm{TaS_2}$  manifest strong Ising-type spin-orbit coupling \cite{delaBarrera2018} and its interplay with helical magnetic order requires a separate consideration.  

\section{s-wave superconductivity suppression in SC/HM heterostructures}
\label{sec:op}

The superconducting order parameter (OP) is to be calculated self-consistently. In terms of the introduced above Green's function the self-consistency equation takes the form:
\begin{align}
   \Delta = \frac{\lambda T}{2} \sum\limits_{\omega_m} \int\frac{d^2p}{(2\pi)^2} {\rm Tr}[\check {{\tilde G}}^{11} \sigma_0 \tau_- \rho_\uparrow]
\label{eq:self-consistency}
\end{align}

\begin{figure}[t]
\includegraphics[width=0.8\columnwidth]{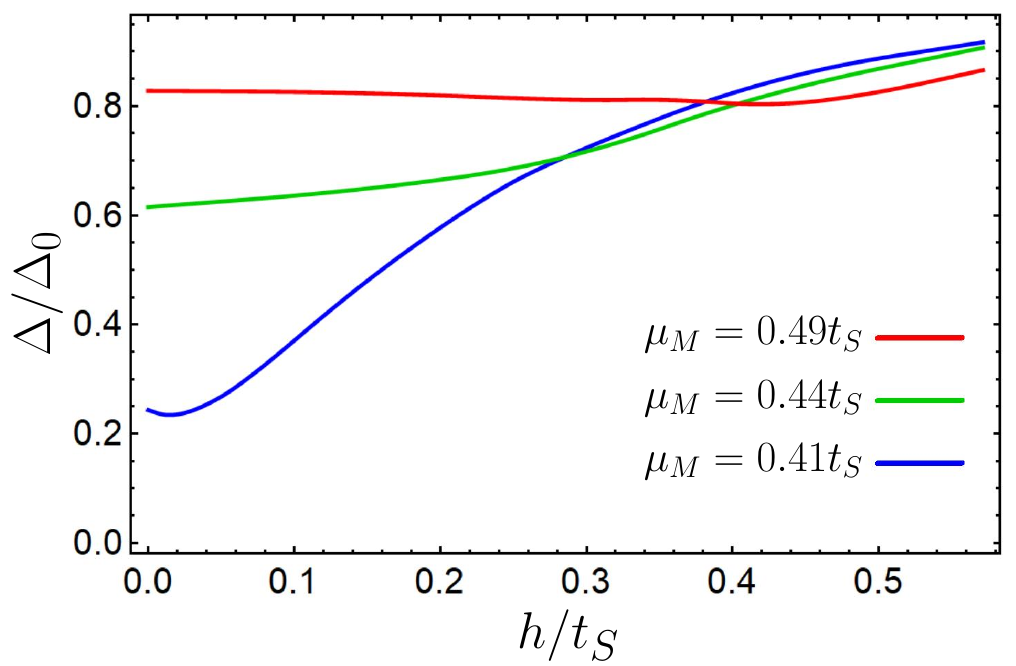}
\caption{Dependence of the superconducting order parameter in SC/HM bilayer on the helical exchange field $h_\perp = h$ of the HM  layer. The magnetization component $h_x$ along the helix axis is assumed to be zero. $\Delta_0$ is the superconducting order parameter of the isolated SC film at temperature $T$. $t_S=70\Delta_0$, $t_M=1.25t_S$, $\mu_S=0.33t_S$, $T=0.35\Delta_0$, $t_{SM}=1.2\Delta_0$.} 
 \label{fig:Delta_h}
\end{figure}

\begin{figure}[t]
\includegraphics[width=0.8\columnwidth]{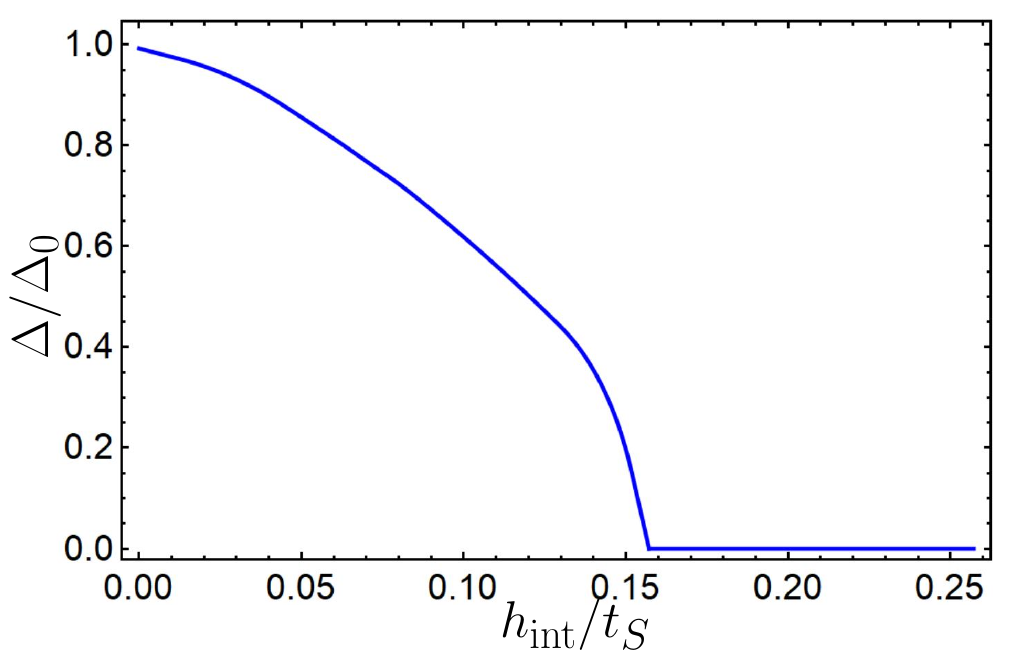}
\caption{Dependence of the superconducting order parameter on the internal helical exchange field $h_{\rm{int}}$ calculated in the framework of the effective homogeneous SC+HM model. The parameters of the hopping hamiltonian describing the SC+HM model correspond to the parameters of the superconducting layer of the SC/HM heterostructure from Fig.~\ref{fig:Delta_h} $t_{SC+HM} = t_S = 70 \Delta_0$, $\mu_{SC+HM} = \mu_S = 0.33 t_S$, $T=0.35\Delta_0$. Parameters $t_M$, $\mu_M$ and $t_{SM}$ are not applicable in this case.} 
 \label{fig:Delta_h_int}
\end{figure}

Here we focus on the helical magnetic texture with no constant $x$-component of the magnetization along the helix axis. The dependence of the OP in the SC/HM bilayer on the helix exchange field magnitude $h_\perp \equiv h$ is shown in Fig.~\ref{fig:Delta_h} for several different values of $\mu_M$. For comparison in Fig.~\ref{fig:Delta_h_int} the analogous dependence of the OP on the internal helical exchange field $h_{\rm{int}}$ calculated in the framework of SC+HM model is demonstrated. It is evident that the dependencies shown in Figs.~\ref{fig:Delta_h} and \ref{fig:Delta_h_int} exhibit the opposite character. In the SC+HM model, OP is monotonically suppressed by the field, although the full suppression occurs at very high exchange fields, significantly exceeding the paramagnetic Pauli limit $\Delta_0/\sqrt 2$  \cite{Sarma1963,Maki1964}, as it was predicted in Ref.~\cite{Sukhachov2025}. At the same time, the calculation for the heterostructure gives a non-monotonic dependence of the OP on the exchange field with an increase at large exchange fields.

\begin{figure}[t]
\includegraphics[width=0.8\columnwidth]{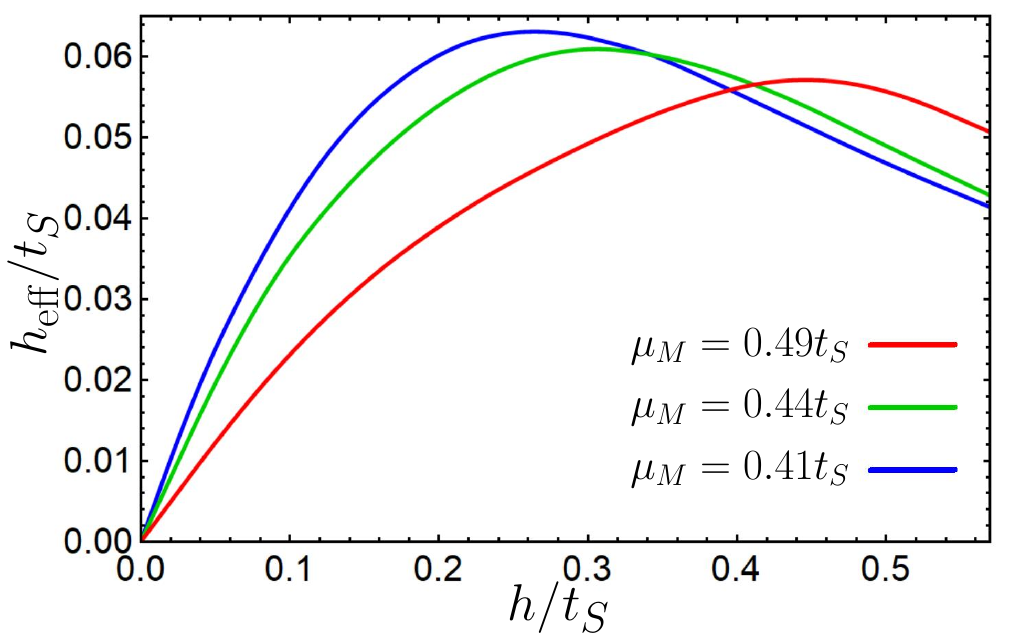}
\caption{Dependence of the effective exchange field $h_{\rm{eff}}$, which is appropriate for description of the the SC/HM heterostructure in the framework  of the SC+HM model on the true
exchange field $h$ in the magnetic layer of the SC/HM heterostructure. Parameters are the same as in Fig.~\ref{fig:Delta_h}.} 
 \label{fig:h_eff}
\end{figure}

Do the opposite trends seen in Figs.~\ref{fig:Delta_h} and \ref{fig:Delta_h_int} mean that SC+HM is not applicable to describe the proximity effect in SC/HM heterostructures? The answer to this question is negative. In fact, it is applicable, one just need to take into account that the exchange field that we introduce into this model is an effective quantity that strongly depends on the difference in the Fermi momenta of the superconductor and the magnet (interface mismatch) and the interlayer hopping parameter (Fermi surface barrier). The dependence of the effective exchange field $h_{\rm{eff}}$ on the true exchange field $h$ in the SC/HM heterostructure is plotted in Fig.~\ref{fig:h_eff}. $h_{\rm{eff}}$ is calculated from the considerations that the same values of triplet superconducting correlations are obtained in the SC/HM heterostructure and in the effective SC+HM model describing the heterostructure, see Sec.~\ref{sec:triplets} for more details. Two key features can be noted. First of all, $h_{\rm{eff}}$ is a nonmonotonic function of $h$, what partially explains the recovering of the OP at high $h$. Second, the amplitude of $h_{\rm{eff}}$ is by the order of magnitude smaller than the true exchange field in the HM. 

It is rather difficult to write down a relatively simple analytical expression describing the behavior of $h_{\rm{eff}}$ for the SC/HM heterostructure, but qualitatively it behaves as $h_{\rm{eff}} \sim h(t_{SM}^2/\Delta E_{SM}^2)$, where $\Delta E_{SM} = \varepsilon_S - \varepsilon_M$ is the difference between the electronic energies of the magnet and the superconductor (superconductor should be considered in its normal state) at a given momentum in the vicinity of the Fermi surface, which is directly related to the mismatch between the Fermi surfaces of the magnet and the superconductor. This expression explains mentioned above key features of  $h_{\rm{eff}}$. First of all, $h_{\rm{eff}}$ is nonmonotonic because  its initial growth at small $h$ is changed by decrease at large $h$ caused by the mismatch between the Fermi surfaces, which inevitably grows at large enough exchange splitting field of the magnet. Second, the above qualitative expression is perturbative and is only valid at  $t_{SM}^2/\Delta E_{SM}^2 \ll 1$. At $\Delta E_{SM} \to 0$ $h_{\rm{eff}}$ reaches its maximum $h_{\rm{eff}} \sim h$.  In general situation the mismatch of the Fermi surfaces of the materials is essential and condition $t_{SM}^2/\Delta E_{SM}^2 \ll 1$ is fulfilled to a good accuracy. For this reason typically $h_{\rm{eff}} \ll h$, as it is demonstrated in Fig.~\ref{fig:h_eff}. Blue curve in Fig.~\ref{fig:h_eff} corresponds to the maximal possible value of $h_{\rm{eff}}$  for given hopping parameters of the HM and SC materials. $h_{\rm{eff}}$ can reach its maximal value $h_{\rm{eff}} \sim h$ for an ideal situation of no mismatch, but in this case the superconductivity is fully suppressed by another depairing factor - leakage of the superconducting correlations to the magnet, which in the framework of the effective SC+HM model is to be described by the Dynes parameter $\Gamma \sim \Delta_0 t_{SM}^2/\Delta E_{SM}^2$ (if one calculates the OP for the SC/HM heterostructure without changing it by the SC+HM model, there is no need to introduce this artifical parameter, the leakage is taken into account automatically). The superconductivity is already fully suppressed at $\Gamma$ considerably smaller than $\Delta_0$. According to the arguments presented above it seems hardly possible to achieve high values of $h_{\rm{eff}} \gg \Delta_0$ in SC/HM heterostructures and simultaneously to not destroy superconductivity. This is despite the fact that the helical exchange field by itself suppresses superconductivity very weakly, as it was reported in Ref.~\cite{Sukhachov2025}.  

\begin{figure}[t]
\includegraphics[width=0.8\columnwidth]{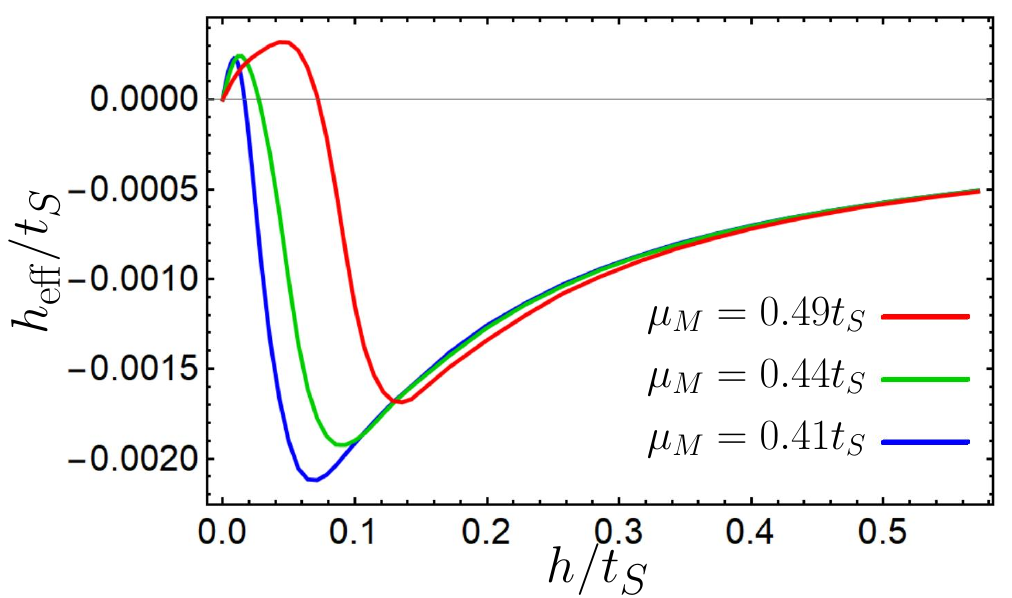}
\caption{Dependence of the effective exchange field $h_{\rm{eff}}$, which is appropriate for description of the the SC/FM heterostructure in the framework  of the SC+FM model on the true
exchange field $h$ of the FM layer. Parameters are the same as in Fig.~\ref{fig:Delta_h}.} 
 \label{fig:h_eff_F}
\end{figure}

The described behavior of $h_{\rm{eff}}$ in SC/HM heterostructures is conceptually similar to its behavior in heterostructures with a ferromagnetic layer having spatially homogeneous, not helical, magnetization, which we will refer to as SC/FM heterostructures \cite{Bobkov2024_hybridization,Ianovskaia2024}. In this case the qualitative estimate $h_{\rm{eff}}\sim h(t_{SM}^2/\Delta E_{SM}^2)$ is also valid. However, for SC/FM heterostructures the expression for $h_{\rm{eff}}$ can be found explicitly and in the framework of the same tight-binding model takes the form \cite{Ianovskaia2024}:
\begin{align}
    h_{{\rm eff}}=\frac{h t_{SM}^2}{h^2-(\mu_S(t_M/t_S)-\mu_M)^2} .
    \label{heff_1_final}
\end{align}
The denominator of this expression just corresponds to $\Delta E_{SM}^2$ at the Fermi level of the superconductor in its normal state. The dependence of  $h_{\rm{eff}}$ on the true exchange field $h$ in the SC/FM heterostructure with the same numerical parameters as for the SC/HM heterostructure is plotted in Fig.~\ref{fig:h_eff_F}. The same general trends -nonmonotonic behavior and the small amplitude of $h_{\rm{eff}}$ are also seen. However, it is worth noting that in contrast to the helical magnetism, in SC/FM heterostructures with the homogeneous ferromagnetic order the superconductivity is fully suppressed by much smaller $h_{\rm{eff}}$ due to the existence of the paramagnetic Pauli limit $h_{\rm{eff}} = \Delta_0/\sqrt 2$ \cite{Clogston1962,Chandrasekhar1962,Sarma1963}. 

In fact, the main depairing factor in the SC/HM heterostructure is the leakage of the superconducting correlations into the HM region. Indeed, according to the data presented in Fig.~\ref{fig:Delta_h_int}, the suppression of the OP in the SC+HM model is very weak at $h_{\rm{int}}$ of the order of typical $h_{\rm{eff}}$ in Fig.~\ref{fig:h_eff}. At the same time, the OP suppression for the SC/HM heterostructure at the same parameters (blue curve in Fig.~\ref{fig:Delta_h}) is not small and cannot be only ascribed to the effect of $h_{\rm{eff}}$. By the order of magnitude this suppression is the same as in a superconductor/normal metal heterostructures and in the framework of the effective SC+HM model can be taken into account by introducing the depairing Dynes parameter $\Gamma$. However, this approach should be used with great caution, since, as discussed above, the Dynes parameter itself in this case  depends very strongly on the parameters of the materials, in particular $h$ and $\mu_{S,M}$.

\begin{figure}[t]
\includegraphics[width=0.8\columnwidth]{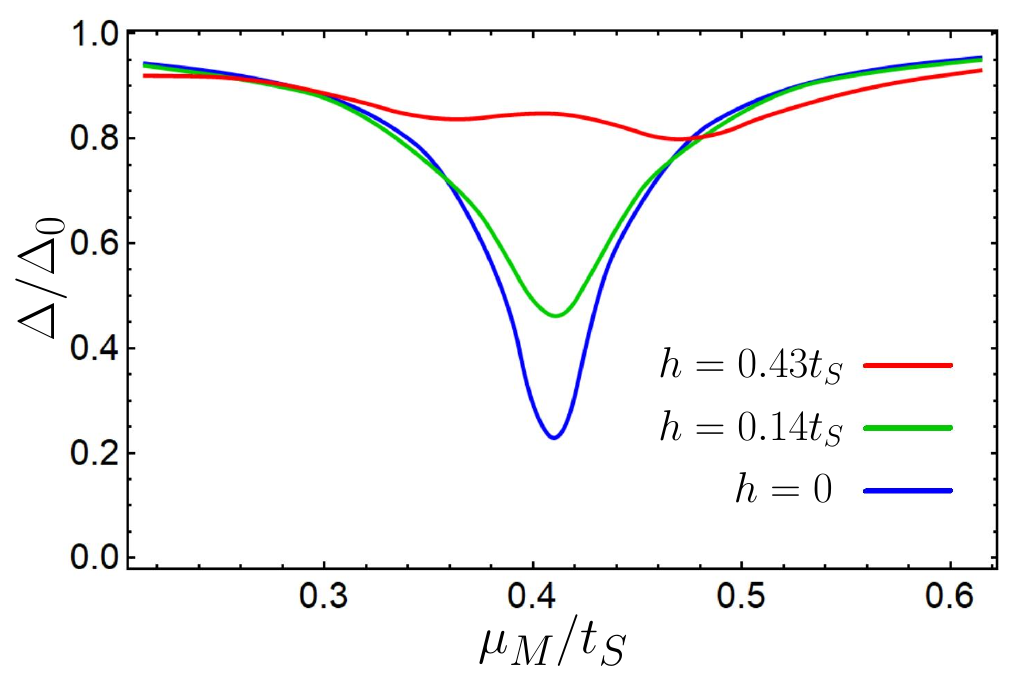}
\caption{Dependence of the superconducting order parameter on the helimagnet on-site energy $\mu_M$ for several values of the helimagnet exchange field. $h_x=0$, $h=0$ corresponds to the case of the nonmagnetic layer in proximity to the SC layer. The other parameters are the same as in Fig.~\ref{fig:Delta_h}.} 
 \label{fig:Delta_mu}
\end{figure}

The physical picture discussed above is further supported by the results of calculating the dependence of the OP on the HM on-site energy $\mu_M$ for different values of $h$, which are shown in Fig.~\ref{fig:Delta_mu}. It is seen that in a wide range of $\mu_M$ values the strongest suppression of the OP occurs at $h=0$. The reason is that $h_{\rm{eff}}$ is small, as discussed above, and the main role of the true exchange field $h$ of the HM is not to suppress superconductivity by the direct depairing via spin splitting, but is to change the mismatch between the SC and HM electronic spectra.

For conical magnets the nonzero homogeneous $x$-component of the effective exchange $h_x \neq 0$ makes the system more similar to the superconductor/ferromagnet heterostructure, where the superconducting OP is fully suppressed at $h_{\rm{eff}} \sim \Delta_0$ and superconducting state becomes energetically unstable at the paramagnetic limit $h_{\rm{eff}} = \Delta_0/\sqrt 2$. Thus, with increasing $h_x$ the OP suppression becomes stronger.

\section{Proximity-induced $P$-wave triplets in SC/HM heterostructures}
\label{sec:triplets}

Now we investigate triplet superconducting correlations, which are induced in the SC/HM heterostructure due to the proximity effect between SC and HM layers.  The triplet correlations are encoded in the anomalous Green's function $F(\bm p)$, which is a matrix in spin, sublattice and layer spaces. In spin space it can be expanded over Pauli matrices:
\begin{align}
    F(\bm p) = F_s (\bm p) \sigma_0 + \sum \limits_{i=x,y,z} F_t^i (\bm p) \sigma_i ,
    \label{eq:anomalous_spin}
\end{align}
where $F_s$ is the singlet component of the anomalous Green's function and $\bm d(\bm p) = (F_t^x, F_t^y, F_t^z)^T$ is the $\bm d$-vector describing triplet correlations. $F_t^{y,z}$ components of the $\bm d$-vector lie in the plane of the helical magnetization and their local in the sublattice space components $F_t^{y,z; kk}$ have the property $\sum \limits_{k=1}^4 F_t^{y(z); kk} = 0$ due to the symmetry reasons, because the helical component of the HM exchange field $\bm h_\perp$ has this property. For this reason these components are not of interest for us further because we are interested in the dissipationless $\it transport$ spin current, which is directly determined by the triplet component $F_t^x$, as it is shown in Sec.~\ref{sec:current}. Since the source of $F_t^x$ is the nonzero cross product $\bm h_{i1} \times \bm h_{i2}$, its amplitude scales as $h_\perp^2$ at small exchange fields and does not change sign in the unit cell. Therefore, further we only discuss $F_t^x$. 

\subsection{SC/HM heterostructure vs a superconductor with helical magnetization}

\begin{figure}[t]
\includegraphics[width=0.8\columnwidth]{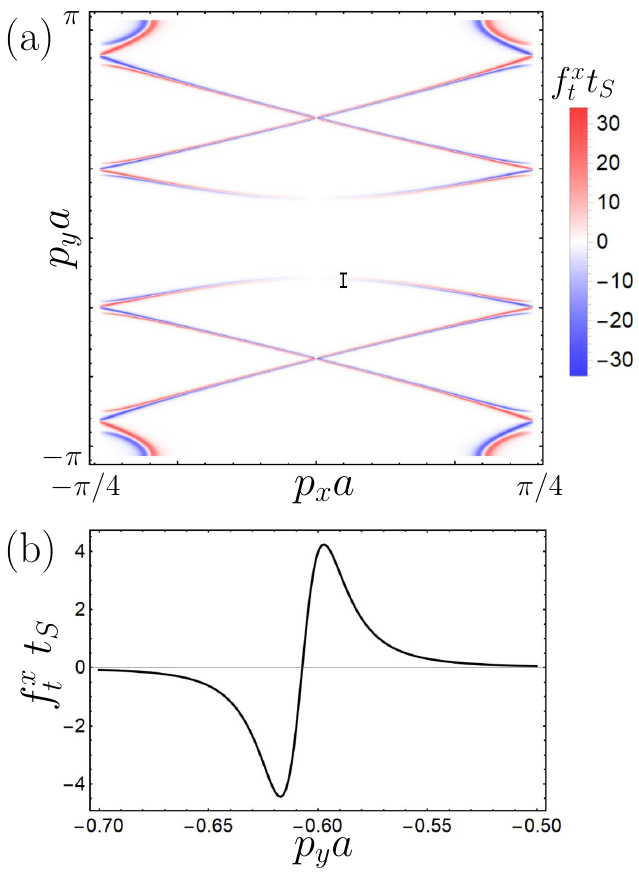}
\caption{SC+HM model of the homogeneous helimagnetic superconductor. (a) Local triplet $p$-wave correlations in the whole Brillouin zone. (b)  Momentum dependence of the local triplet $p$-wave correlations along the short black line in panel (a). $h_{\rm{int}}=0.14t_S$, $\Delta=0.8\Delta_0$. The other parameters are the same as in Fig.~\ref{fig:Delta_h_int}.} 
 \label{fig:triplets_model}
\end{figure}

\begin{figure}[t]
\includegraphics[width=0.8\columnwidth]{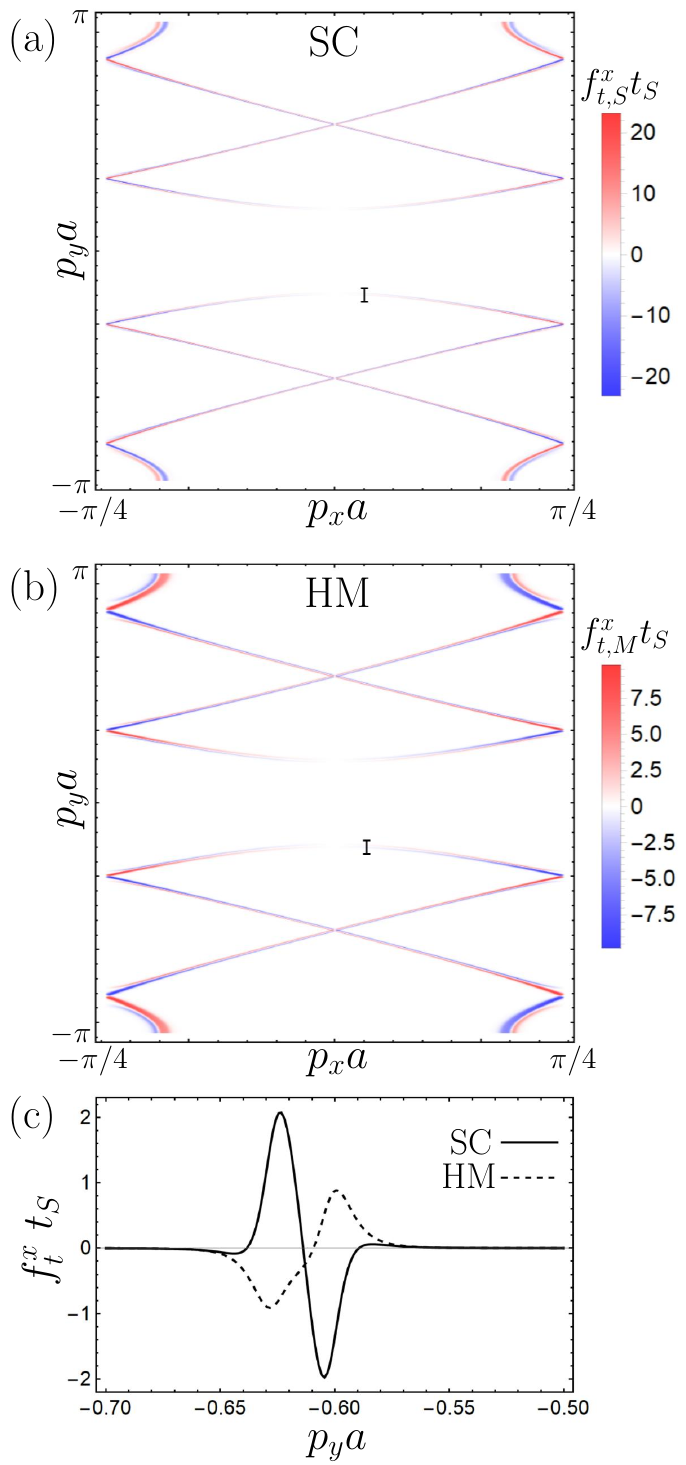}
\caption{SC/HM heterostructure. Local triplet $p$-wave correlations in the whole Brillouin zone (a) in the SC layer and (b) in the HM layer. (c) Momentum dependence of the local triplet $p$-wave correlations along the short black line in panels (a) and (b).  $h=0.14t_S$, $\mu_M=0.41t_S$, $\Delta=0.8\Delta_0$. The other parameters are the same as in Fig.~\ref{fig:Delta_h}.} 
 \label{fig:triplets_heterostructure}
\end{figure}

First of all, we focus on the case of a helical magnet with $h_x=0$ and compare the $p$-wave triplet correlations in SC/HM heterostructure and in the SC+HM model. Local component of $F_t^x$, which is defined as $f_t^x = \sum \limits_{k=1}^4 F_t^{x;kk}$, calculated for the homogeneous superconductor in the presence of the helical exchange field (SC+HM model) is demonstrated in Fig.~\ref{fig:triplets_model}(a) in the whole Brillouin zone (BZ). The BZ is compressed $4$ times along the $p_x$-axis due to the fact that the unit cell contains 4 sites along the $x$-direction. $a$ is the lattice constant. The amplitude of $f_t^x$ differs significantly from zero only in the vicinity of Fermi surfaces, which is absolutely standard behavior for superconducting correlations of any type due to the smallness of the binding energy of the Cooper pair. $f_t^x$ demonstrates the odd-parity behavior $f_t^x(\bm p) = -f_t^x(-\bm p)$. Analogous $p$-wave behavior of the triplet correlations in $s$-wave superconductors with $p$-wave magnetism was recently reported \cite{Maeda2025}. Taking into account that a helimagnetic exchange field is equivalent to the $p$-wave magnetic texture in the presence of a homogeneous exchange field \cite{Martin2012}, our result is in agreement with the result of Ref.~\cite{Maeda2025}. If we look more closely at the structure of $f_t^x$ in the vicinity of the Fermi surface, it turns out that with good accuracy it is antisymmetric  relative to the Fermi surface. To see it more clearly, Fig.~\ref{fig:triplets_model}(b) shows $f_t^x$ along a segment of the Brillouin zone, marked by a short black line in Fig.~\ref{fig:triplets_model}(a). This shape of the $f_t^x$ anomalous Green's function describing $p$-wave triplet correlations is very different from the behavior of singlet correlations and $F_t^{y,z}$ triplet correlations, which have nearly symmetric distribution around the Fermi surface. The antisymmetry of $f_t^x$ plays important role in spintronics applications of the SC/HM heterostructures leading to a strong reduction of the dissipationless transport spin current, as it is discussed in Sec.~\ref{sec:current}.

Fig.~\ref{fig:triplets_heterostructure} represents the same $p$-wave superconducting triplet component in SC/HM heterostructure. In this case we have two spatially separated layers and, therefore, we plot the $p$-wave triplet components in these layers separately.  The triplet component induced in SC part of the heterostructure $f_{t,S}^x = \sum \limits_{k=1}^4 {\rm Tr}[F_t^{x;kk}\rho_\uparrow]$ is shown in Fig.~\ref{fig:triplets_heterostructure}(a) , and the triplet component induced in HM part of the heterostructure $f_{t,M}^x = \sum \limits_{k=1}^4 {\rm Tr}[F_t^{x;kk}\rho_\downarrow]$ is shown in Fig.~\ref{fig:triplets_heterostructure}(b). Both figures look qualitatively similar to the picture of $p$-wave triplet correlations in the homogeneous helimagnetic superconductor presented in Fig.~\ref{fig:triplets_model}. The only essential difference is that the average amplitude of the triplet correlations in the SC/HM heterostructure with the helical exchange field $h$ in the HM layer is several times smaller than in the SC+HM model corresponding to the same value of the internal helical exchange field $h_{\rm{int}} = h$. In order to directly compare the amplitudes of the $p$-wave triplet correlations in SC+HM model and in the SC/HM heterostructure with the same value of the true exchange field $h$ we take the same value of $\Delta$ and do not calculate it self-consistently.  

Fig.~\ref{fig:triplets_heterostructure}(c) shows $f_{t,S,M}^x$ along a segment of the Brillouin zone, marked by a short black line in Figs.~\ref{fig:triplets_heterostructure}(a) and (b). It is seen that the shape of $f_{t,S,M}^x$ is qualitatively similar to the SC+HM model, compare to Fig.~\ref{fig:triplets_model}(c). It is interesting that in the HM layer  $f_{t,M}^x$ has the same sign as in the SC+HM model, however in the SC layer the sign of  $f_{t,S}^x$ is reversed. It means that the spin splitting induced in the SC layer by proximity to the HM layer for this particular part of the BZ is opposite to the spin-splitting in the HM layer. Similar effect can take place even for superconductor/homogeneous ferromagnet heterostructures, as it was discussed in Refs.~\cite{Bobkov2024_hybridization,Ianovskaia2024} and also seen from Eq.~(\ref{heff_1_final}).  

Based on the above, we can conclude that from the point of view of the main manifestation of the proximity effect - the induction of triplet correlations, the heterostructure can be described with good accuracy within the framework of a homogeneous SC+HM model, but with some effective internal exchange field $h_{\rm{eff}}$ that gives the correct amplitude of triplet correlations at the same value of $\Delta$. We find $h_{\rm{eff}}$ from the condition that the averaged over the entire BZ $f_t^x$ in the SC+HM model is equal to the averaged over the entire BZ $f_{t,S}$ in the superconducting part of the SC/HM heterostructure. $h_{\rm{eff}}$ calculated according to this method is shown in Fig.~\ref{fig:h_eff}.

\subsection{SC/HM heterostructures with conical magnets}

\begin{figure}[t]
\includegraphics[width=0.8\columnwidth]{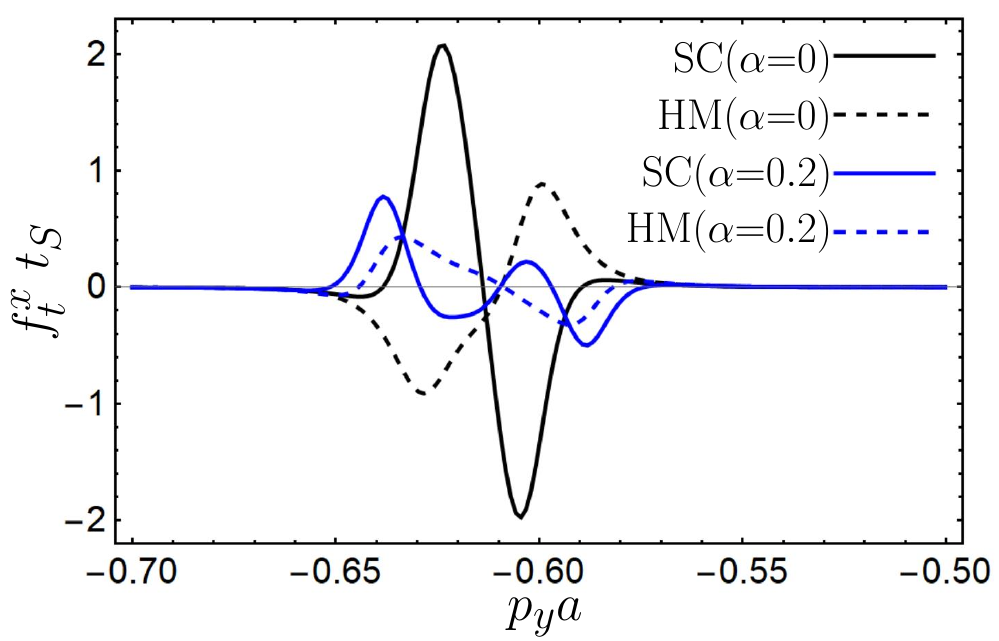}
\caption{Comparison of SC/HM heterostructures with helical and conical magnets. Momentum dependence of the local triplet $p$-wave correlations along the short black lines in Figs.~\ref{fig:triplets_heterostructure}(a) and (b).  The $p$-wave correlation amplitudes for the helical (conical) magnet case $h_x=0$ ($h_x \neq 0$) are shown by black (blue) solid lines for correlations in the SC layer and dashed lines for correlations in the HM layer lines. Parameters are the same as in Fig.~\ref{fig:triplets_heterostructure}.} 
 \label{fig:triplets_conical}
\end{figure}

Here we discuss the influence of the homogeneous $h_x$ component of the spiral exchange field of the HM layer on the $p$-wave triplet correlations. Its value can be quantified by the angle $\alpha$ between $(y,z)$-plane and the HM magnetization, that is $h_x = h \sin \alpha$ and $h_\perp = h \cos \alpha$. In Fig.~\ref{fig:triplets_conical} the shape of $f_{t,S,M}^x$ is compared for the SC/HM heterostructures with helical ($\alpha = 0$) and conical ($\alpha \neq 0$) magnets. As the angle $\alpha$ increases, two processes occur simultaneously. First, the amplitude of $f_{t,S,M}$ decreases. This is natural, since it is proportional to $h_\perp^2$ and vanishes at $\alpha = \pi/2$. But more importantly, as the angle $\alpha$ increases, the correlations lose their antisymmetry relative to the Fermi surface and the degree of asymmetry increases. In the next section, it will be shown that the latter property leads to an increase in the magnitude of the transport superconducting spin current with an increase in the degree of conicity of the magnet. 

\section{Superconducting spin current}
\label{sec:current}

In the framework of the considered nearest-neighbor tight-binding model the spin currentcarrying $\alpha$-spin component from $i\nu$-site to the nearest right neighbor along the $x$-axis can be calculated as 
\begin{align}
    &I_{x,\nu=(1,2,3)}^\alpha=- \frac{i \hbar}{2}\sum_{ s s' } \left \langle\hat c^\dagger_{i\nu s} (\sigma_{\alpha})_{ss'} \hat t\hat c_{i(\nu+1) s'} -h.c.\right \rangle,  \nonumber \\
    &I_{x,\nu=4}^\alpha=-\frac{i \hbar}{2} \sum_{ s s' } \left \langle \hat c^\dagger_{i\nu s} (\sigma_{\alpha})_{ss'} \hat t \hat c_{(i+4a_x)(\nu-3) s'} -h.c.\right \rangle.
    \label{eq:spin_current_def}   
\end{align}
The spin current along the $y$-axis can be written as
\begin{align}
I_{y,\nu}^\alpha=-\frac{i \hbar}{2} \sum_{ s s' } \left \langle \hat c^\dagger_{i\nu s} (\sigma_{\alpha})_{ss'}\hat t \hat c_{(i+a_y)\nu s'} -h.c.\right \rangle .
\label{eq:spin_current_def2}
\end{align}
The electric current takes the form
\begin{align}
    &I_{x,\nu=(1,2,3)}=- i e\sum_{s  } \left \langle \hat c^\dagger_{i\nu s}\hat t \hat c_{i(\nu+1) s} -h.c.\right \rangle,  \nonumber \\
    &I_{x,\nu=4}=-i e \sum_{ s  } \left \langle \hat c^\dagger_{i\nu s} \hat t\hat c_{(i+4a_x)(\nu-3) s} -h.c.\right \rangle , \nonumber \\
    &I_{y,\nu}=-i e \sum_{ s } \left \langle \hat c^\dagger_{i\nu s}\hat t  \hat c_{(i+a_y)\nu s} -h.c.\right \rangle.
\label{eq:electric_current_def}
\end{align}
We have checked that $\sum \limits_{k=1}^4 I_{x(y),\nu}^{y,z} = 0$, as it is dictated by symmetry reasons. At the same time $I_{x(y),\nu}^x \equiv I_{x(y)}^x$ does not depend on the sublattice index $\nu$. Then in terms of the Green's function introduced in Eq.~(\ref{eq:gorkov_mixed}) it can be expressed as
\begin{align}
    I^{x}_{x(y)}=\frac{T\hbar}{16}\sum_{\omega_m}\int \frac{d^2p}{(2\pi)^2} {\rm Tr}[\check {{\tilde G}} \sigma_x (\tau_0+\tau_z) \rho_0 \check M_{x(y)}],
    \label{eq:spin_current_GF}
\end{align}
where
\begin{align}
    \check M_x=-i\hat t \left(
\begin{array}{cccc}
0 &  \kappa & 0 & \kappa e^{4ip_xa} \\
-\kappa^* &  0 & \kappa & 0 \\
0 &  -\kappa^* & 0 & \kappa \\
-\kappa^*e^{-4ip_xa} &  0 & -\kappa^* & 0 
\end{array}
\right)_l 
\end{align}
with $\kappa = e^{iq_xa/2}$.
\begin{align}
    \check M_y=2 \hat t\sin((p_y+q_y/2) a) \left(
\begin{array}{cccc}
1 &  0 & 0 & 0 \\
0 &  1 & 0 & 0 \\
0 &  0 & 1 & 0 \\
0 &  0 & 0 & 1 
\end{array}
\right)_l .
\end{align}
The electric current $I_{x(y)}$ is obtained from Eq.~(\ref{eq:spin_current_GF}) by the substitution $\hbar/2 \to e$ and $\sigma_\alpha \to \sigma_0$.

\begin{figure}[t]
\includegraphics[width=0.8\columnwidth]{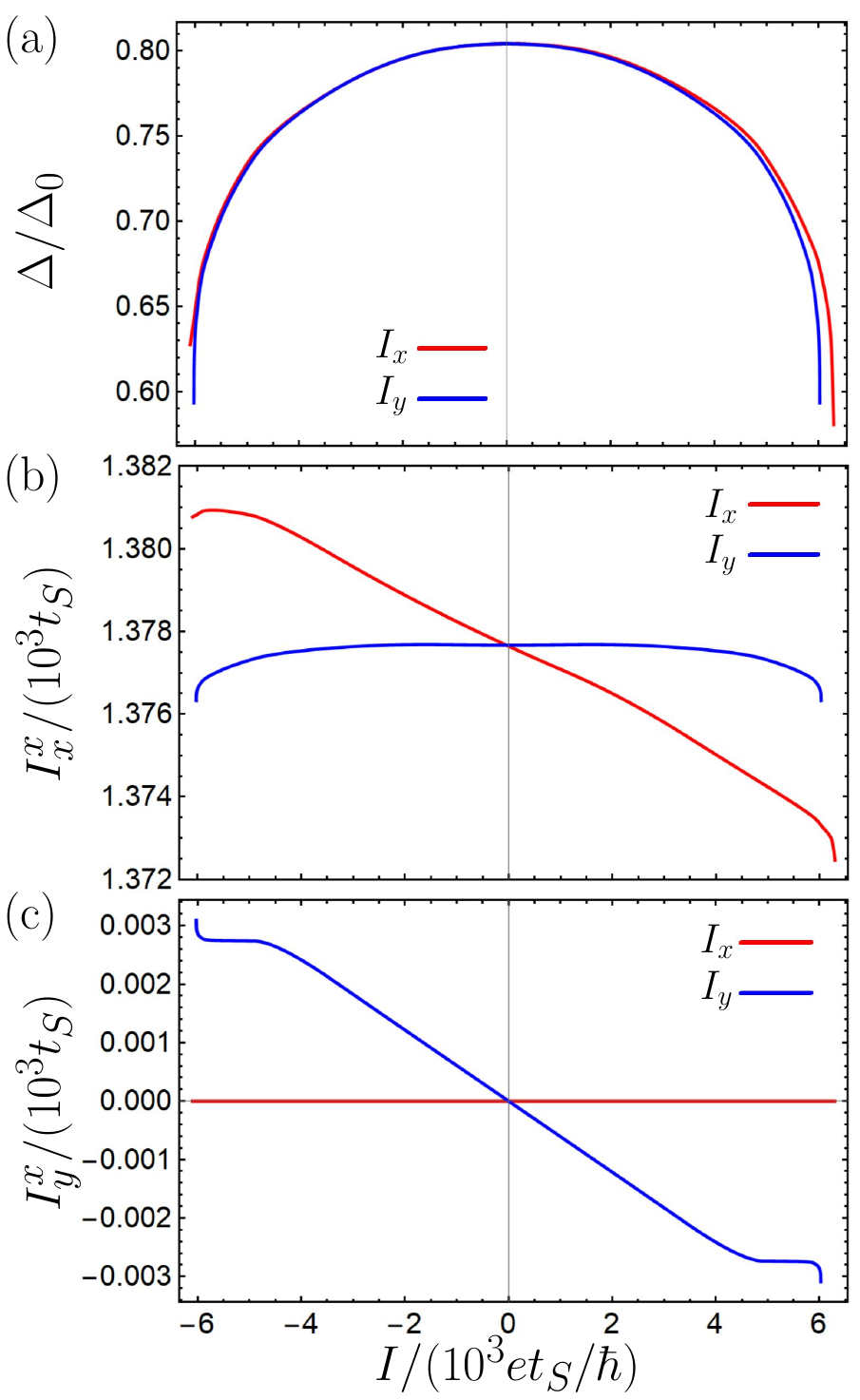}
\caption{(a) Superconducting OP as a function of the applied electric supercurrent. (b)-(c) Spin current carrying $x$-component of spin and flowing along the $x$-axis (b) and along the $y$-axis (c) as a function of the applied electric supercurrent. Blue (red) lines in all panels correspond to the case when the electric supercurrent is applied along the $x(y)$-axis. $h=0.14t_S$, $\mu_M=0.41t_S$, $t_{SM}=0.8\Delta_0$, $\alpha=0.2$, other parameters are the same as in Fig.~\ref{fig:Delta_h}.} 
 \label{fig:spin_current_general}
\end{figure}

In Fig.~\ref{fig:spin_current_general}(b) and (c) $x$-(along the helix axis) and  $y$-(perpendicular to the helix axis) components of the spin current calculated according to Eq.~(\ref{eq:spin_current_GF}) are shown as functions of the electric supercurrent applied to the bilayer. The spin supercurrent is calculated self-consistently and the corresponding dependence of the OP on the applied supercurrent is shown in Fig.~\ref{fig:spin_current_general}(a). It is seen that the critical supercurrent, which fully destroys superconductivity, takes slightly different values if the supercurrent is applied in positive and negative directions of the helix axis ($x$-axis). That is, the SC/HM heterostructure  manifests a weak superconducting diode effect. 

Turning to the analysis of the spin supercurrent, first of all, from  Fig.~\ref{fig:spin_current_general}(b) it can be seen that the system sustains rather high value of a spontaneous spin current along the helix axis (spontaneous spin supercurrent occurs at zero applied supercurrent $I_x = 0$). This spontaneous spin current is well-known even for non-superconducting systems with spin textures or in the presence of spin-orbit interaction \cite{Sonin1982,Konig2001,Rashba2003} and is not associated with real spin
transport \cite{Rashba2003}.

Application of an electric supercurrent is accompanied by a transport spin supercurrent. This transport spin supercurrent is associated with the $p$-wave triplet Cooper pairs, what is discussed in details below. The nearly linear dependence between the spin and the electric supercurrents is only violated at a high supercurrent close to its critical value. This is because the $I_{x(y)}(\bm q)$ has maximum at the pair momentum $\bm q$ corresponding to the critical value of the supercurrent $I_c$, what leads to the vertical derivative of $I_{y}^x(I)$ at $I=I_c$ in Fig.~\ref{fig:spin_current_general}(c). Also, due to the depairing effect of the supercurrent on superconductivity the supercurrent applied perpendicular to the helix axis (along the $y$-axis) leads to a weak change in the spontaneous spin current flowing along the helix axis, see the blue line in Fig.~\ref{fig:spin_current_general}(b).

\begin{figure}[t]
\includegraphics[width=0.8\columnwidth]{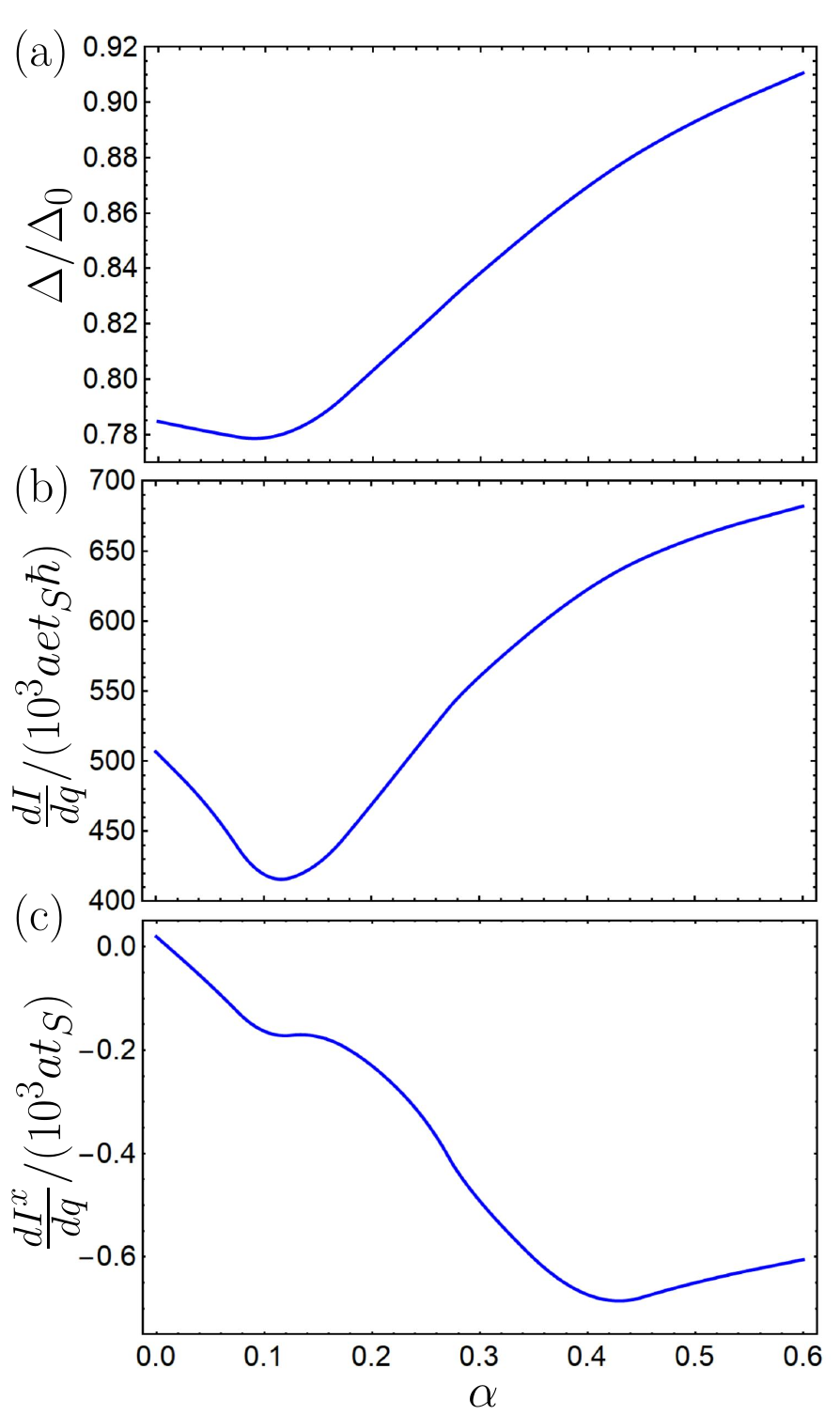}
\caption{SC/HM heterostructure with conical magnet. (a) Superconducting OP as a function of the angle $\alpha$ between $(y,z)$-plane and the HM magnetization. (b) Electric (b) and transport spin (c) supercurrents as a function of $\alpha$. The electric and spin supercurrents are calculated in the linear-order approximation with respect to the pair momentum $\bm q$ and for this reason only their derivatives with respect to $\bm q$ are showm, that is $\bm I = (dI/dq) \bm q$ and $\bm I^x = (dI^x/dq) \bm q$. $h=0.14t_S$, $\mu_M=0.41t_S$, $t_{SM}=0.8\Delta_0$, other parameters are the same as in Fig.~\ref{fig:Delta_h}.} 
 \label{fig:spin_current_conical}
\end{figure}

To demonstrate that the transport spin supercurrent is carried by $p$-wave triplet correlations, we expand the full expression for the current Eq.~(\ref{eq:spin_current_def2}) to  first order in the Cooper pair momentum $\bm q$. The zeroth-order term determines the spin spontaneous current, and the first-order term determines the transport spin current. Then the Green's function can be taken in the form
\begin{align}
    \check {\tilde G}=\check {\tilde G}_0+\delta \check {\tilde G} ,
\end{align}
where $\check {\tilde G}_0 = \check {\tilde G} (\bm q = 0)$ and $\delta \check {\tilde G}$ is the first-order correction. From Eq.~(\ref{eq:gorkov_mixed}) it can be found that 
\begin{align}
    \delta \check {\tilde G}=-\left(
\begin{array}{cc}
G_0 &  F_0  \\
\bar F_0 &  \bar G_0 
\end{array}
\right) \frac{d\hat \xi}{d\bm p} \frac{\bm q}{2} \left(
\begin{array}{cc}
0 &  F_0  \\
-\bar F_0 &  0 
\end{array}
\right).
\end{align}
The normal part of the first-order correction takes the form:
\begin{align}
    \delta G=F_0 \frac{d\hat \xi}{d\bm p} \frac{\bm q}{2} \bar F_0 .
\end{align}
The $x$-component of $\delta G$ in spin space, which contributes to $I_{x(y)}^x$ according to Eq.~(\ref{eq:spin_current_GF}), is expressed in terms of different spin components of the anomalous Green's function as follows:
\begin{align}
    \delta G^x=F_{0t}^x \frac{d\hat \xi}{d\bm p} \frac{\bm q}{2} \bar F_{0s}+F_{0s} \frac{d\hat \xi}{d\bm p} \frac{\bm q}{2} \bar F_{0t}^x+  \nonumber \\
    iF_{0t}^y \frac{d\hat \xi}{d\bm p} \frac{\bm q}{2} \bar F_{0t}^z-iF_{0t}^z \frac{d\hat \xi}{d\bm p} \frac{\bm q}{2} \bar F_{0t}^y .
    \label{eq:delta_G_x}
\end{align}
After summation over Matsubara frequencies and integration over $\bm p$ the contribution of the second line of Eq.~(\ref{eq:delta_G_x}) to the superconducting spin current Eq.~(\ref{eq:spin_current_GF}) vanishes. Then the spin current is determined by the $p$-wave triplet correlations discussed in Sec.~\ref{sec:triplets}. The integration over $d^2p/(2 \pi)^2$ can be represented as integration over the momentum components perpendicular and parallel to the Fermi surface. As it was shown in Sec.~\ref{sec:triplets}, at $h_x = 0$ $(\alpha =0)$ the $p$-wave triplet correlations are nearly antisymmetric around the Fermi surface along the perpendicular direction, and the singlet correlations are nearly symmetric. For this reason the spin current is strongly suppressed at $\alpha = 0$. With increasing $\alpha$ the degree of asymmetry of the $p$-wave triplets increases, which leads to a significant increase in the magnitude of the spin current. 

The transport spin currents $I_{x}^x(I_x)$  and $I_{y}^x (I_y)$ calculated in the linear approximation with respect to the pair momentum $\bm q$ coincide with high accuracy and are shown in Fig.~\ref{fig:spin_current_conical}(c) as functions of $\alpha$. It is evident that the amplitude of the spin current grows on average approximately linearly with increasing $\alpha$ up to values of the order of $\alpha \sim 0.4$. Then the amplitude of the spin current decreases. The reason is that with a further increase of $\alpha$ the magnet becomes closer to a homogeneous ferromagnet than to a helical magnet and, accordingly, the amplitude of the $p$-wave triplets drops. Therefore, in contrast to heterostructures with helical magnets  and ferromagnets SC/HM heterostructures with conical magnets are capable of transferring dissipationless spin current.

The other specific feature of the spin current at $\alpha \sim 0.1$ is correlated with the minimum of the OP at this value of $\alpha$, see Fig.~\ref{fig:spin_current_conical}(a). This minimum is the result of a complicated dependence of $h_{\rm{eff}}$ and $\Gamma$ on $\alpha$. The local minimum of the OP leads to the local suppression of the electric current [Fig.~\ref{fig:spin_current_conical}(b)] and the spin current [Fig.~\ref{fig:spin_current_conical}(c)] because they both are carried by the Cooper pairs.  The growth of the OP and superconducting current as the magnet approaches the ferromagnetic state seems at first glance to be a counterintuitive result, since it is known that the helical component of the exchange field is effectively averaged and does not exert a large suppressive effect on superconductivity \cite{Sukhachov2025}, while the ferromagnetic component, on the contrary, significantly suppresses singlet superconductivity \cite{Sarma1963}. These intuitive expectations are indeed realized in the SC+HM model, but the physics of the proximity effect in the SC/HM heterostructure turns out to be more complex. 
 
\section{Conclusions}
\label{sec:conclusions}

The proximity effect in superconductor/helical (conical) magnet heterostructures is theoretically studied. It is shown that the SC/HM heterostructure, in principle, can be described in terms of an effective model of a homogeneous superconductor in a homogeneous helical (conical) exchange field (SC+HM model). However, for an adequate description of the proximity effect in the SC/HM heterostructure, the parameters of the effective internal exchange field $h_{\rm{eff}}$ in this model and the Dynes parameter responsible for the decoupling of Cooper pairs must be complex functions of the true exchange field of the magnet and the filling factors of the electron bands of the magnet and superconductor. It can lead to counterintuitive nonmonotonic behavior of the superconducting order parameter and the electric supercurrent in dependence on the amplitude of exchange field of the magnet and on the angle between the exchange field and the helix axis, which determines the degree of the magnet conicity. 

It is predicted that the proximity effect in SC/HM heterostructure induces even-frequency spin-triplet $p$-wave superconducting correlations. The internal structure of the  $p$-wave superconducting correlations in the whole BZ is investigated. It is shown that due to the presence of the $p$-wave superconducting correlations the SC/HM heterostructures with conical magnets are capable of carrying the superconducting spin current. In contrast, the capability of SC/HM heterostructures with helical magnets to carry the spin supercurrent is negligible because of nearly antisymmetric structure of the $p$-wave correlations in the BZ with respect to the Fermi surface, which is violated for conical magnets. The amplitude of the spin supercurrent is dictated by the amplitude of the proximity-induced  $p$-wave triplet component, which in its turn is determined by the amplitude of $h_{\rm{eff}}$. 

The suppression of superconductivity by the helical exchange field is very weak and based on this fact one might expect that it is possible to achieve very large values of $p$-wave correlations and superconducting spin current in SC/HM heterostructures with strong conical magnets. However, our analysis shows that this is not the case. The amplitude of $h_{\rm{eff}}$ is strongly restricted by the mismatch between the magnet and superconductor Fermi surfaces and cannot be done large enough without the full suppression of superconductivity by leakage of Cooper pairs into the HM region.

Thus, we have shown that SC/HM heterostructures with conical magnets are promising objects for superconducting spintronics from the point of view of generating  dissipationless transport spin currents, but the maximum values of spin currents that can be achieved in such systems are inevitably limited. From this point of view, ferromagnetic superconductors with a spiral magnetic order may be more promising candidates, but the advantage of the SC/HM heterostructures is their larger degree of controllability by external parameters, such as gate voltage.

\begin{acknowledgments}
The financial support from the Russian
Science Foundation via the Project No.24-22-00186 is acknowledged.  
\end{acknowledgments}

\bibliography{vdW_spiral}

\end{document}